\documentclass[prl,aps,twocolumn,superscriptaddress,floatfix,showpacs]{revtex4}
\usepackage{graphicx}
\usepackage{color}

\begin{document}

\title{Quantum order by disorder and accidental soft mode in ${\rm\bf Er_2Ti_2O_7}$}

\author{M. E. Zhitomirsky}
\affiliation{Service de Physique Statistique, Magn\'etisme et Supraconductivit\'e,
UMR-E9001 CEA-INAC/UJF, 17 rue des Martyrs, 38054 Grenoble Cedex 9, France}
\author{M. V. Gvozdikova}
\affiliation{Service de Physique Statistique, Magn\'etisme et Supraconductivit\'e,
UMR-E9001 CEA-INAC/UJF, 17 rue des Martyrs, 38054 Grenoble Cedex 9, France}
\author{P. C. W. Holdsworth}
\affiliation{Laboratoire de Physique, \'Ecole Normale Sup\'erieure de Lyon,
CNRS 69364 Lyon Cedex 07, France}
\author{R. Moessner}
\affiliation{Max-Planck-Institut f\"ur Physik komplexer Systeme, 01187 Dresden, Germany}
\date{\today}

\begin{abstract}
Motivated by recent neutron scattering experiments, we derive and study
an effective ``pseudo-dipolar'' spin-1/2 model for the $XY$  pyrochlore antiferromagnet
$\rm Er_2Ti_2O_7$. While a bond-dependent in-plane exchange anisotropy removes
any continuous symmetry, it does lead to a one-parameter `accidental' classical
degeneracy. This degeneracy is lifted by quantum fluctuations in favor of the
noncoplanar spin structure observed experimentally---a rare experimental instance
of quantum order by disorder. A non-Goldstone low-energy mode is present in
the excitation spectrum in accordance with inelastic neutron scattering data.
Our theory also resolves the puzzle of the experimentally observed
continuous ordering transition, absent from previous models.
\end{abstract}
\pacs{75.10.Jm, 
      75.30.Ds, 
      75.50.Ee, 
}
\maketitle

{\it Introduction.}---%
Exotic low-temperature properties of magnetic pyrochlore oxides have their origin
in the underlying frustrated lattice of corner-sharing tetrahedra, which is combined
with diverse crystal-field effects for rare-earth ions \cite{Gardner10}.
Initially, the attention was focused on to spin-ice materials $\rm Ho_2Ti_2O_7$
and $\rm Dy_2Ti_2O_7$ with Ising-like moments \cite{Harris97,Moessner98} and
on Heisenberg antiferromagnets $\rm Gd_2Ti_2O_7$ and $\rm Gd_2Sn_2O_7$ with large
$S=7/2$ spins \cite{Raju99,Wills06}. The above pyrochlore compounds exhibit
various kinds of ordered and unusual disordered classical magnetic phases naturally
expected in view of large rare-earth moments. However, it was gradually
recognized that collective behavior of low-energy magnetic doublets selected by
the local crystalline electric field (CEF) for other rare-earth pyrochlores should be
generally described by effective non-Ising pseudo-spin-1/2 Hamiltonians implying
significant quantum effects \cite{Molavian07,Onoda10,Ross11,Savary12}.

\begin{figure}[tcb]
\centerline{
\includegraphics[width=0.9\columnwidth]{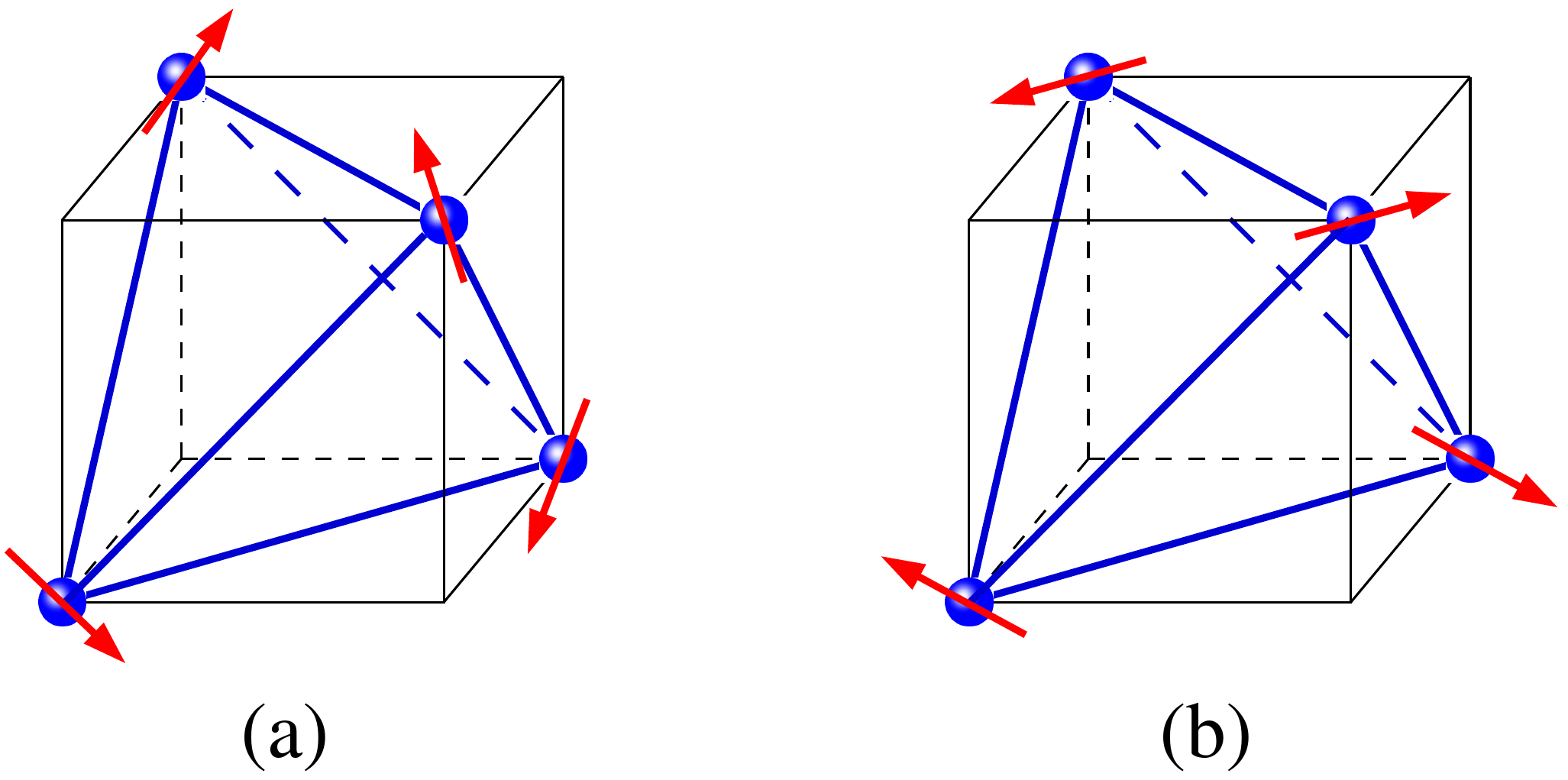}}
\caption{(Color online) Possible $q=0$ spin structures of the $XY$ pyrochlore
antiferromagnet in zero applied field: $\psi_2$ state (a) and $\psi_3$ state (b).
}
\label{fig:Mstates}
\end{figure}

$\rm Er_2Ti_2O_7$ has a special place in the family of magnetic pyrochlores.
It is the only pyrochlore antiferromagnet found so far that exhibits  strong easy-plane
anisotropy with local moments confined to planes orthogonal to the local
$\langle 111\rangle$-axes. Erbium Titanate has been extensively studied in the past
and is well characterized experimentally
\cite{Blote69,Bramwell00,Champion03,Poole07,Ruff08,Sosin10,Cao10,Petrenko11}.
It orders below $T_c\simeq 1.2$~K,  into an unusual $k=0$ non-coplanar antiferromagnetic
structure shown in Fig.~\ref{fig:Mstates}(a), the so called $\psi_2$ state
\cite{Champion03,Poole07}.
This is rather surprising given that the geometry of the four local easy planes
also allows coplanar spin arrangements such as the the $\psi_3$ state shown in
Fig.~\ref{fig:Mstates}(b), with the same energy as the noncoplanar state.
Moreover,  states obtained by {\em reversing} a pair of antiparallel spins in
Fig.~\ref{fig:Mstates}(b) minimize the dipolar energy \cite{Palmer00}.

Quantum order by disorder effect \cite{Shender82} was suggested as selection mechanism 
for the noncoplanar spin structure \cite{Champion03}, because similar thermal order
by disorder selection \cite{Henley89} indeed occurs in a model of classical $XY$ moments
\cite{Bramwell94,Champion03,Champion04,Stasiak11}.
However, the concomitant transition turns out to be first order, in clear contradiction with the
continuous second order transition observed in $\rm Er_2Ti_2O_7$
\cite{Blote69,Champion03}.
Another recent experimental result, which has so far remained unexplained,
is the presence of a quasi-acoustic mode in this highly
anisotropic antiferromagnet found by inelastic neutron scattering (INS)
measurements \cite{Ruff08}.

In this paper we formulate a minimal spin model for $\rm Er_2Ti_2O_7$ which
accounts for this entire body of experimental results: quantum order by disorder
stabilizes the observed ground state, with the low-energy mode in INS a remnant
of the lifted degeneracy; while classical Monte Carlo simulations reproduce
the continuous transition. Using this model, we analyze theoretically various
physical effects such as the energetic selection due to quantum fluctuations,
spin-wave results for the excitation spectrum and the nature of the
finite-temperature phase transition studied by classical Monte Carlo simulations.
We present various estimates of parameters in the model Hamiltonian, which has
analogous form---but opposite sign!---to a dipolar interaction.

{\it Model.}---%
Free $\rm Er^{3+}$ ions ($J=15/2$, $g_J=6/5$) have large magnetic
moments $\mu=g_J\mu_B\sqrt{J(J+1)}= 9.58\mu_B$.
A somewhat reduced value  $\mu=8.9\mu_B$ found in $\rm Er_2Ti_2O_7$
from high-temperature susceptibility measurements \cite{Bramwell00}
reflects a partial quenching of the angular momentum by CEF.
According to crystal-field analysis \cite{Champion03,Dasgupta06},
the ground-state Kramer's doublet $|\pm\rangle$ of $\rm Er^{3+}$ ions
 is strongly anisotropic:
\begin{equation}
\langle \pm|J^z |\pm\rangle = \pm\textstyle\frac{1}{2} \lambda_z,
\quad
\langle -|J^- |+\rangle = \langle +|J^+ |-\rangle = \lambda_\perp.
\label{matrixE}
\end{equation}
with $\lambda_\perp = 6.35$ and $\lambda_z = 0.202$. Neglecting higher
CEF levels, one can use the operator equivalence
\begin{equation}
J^z = \lambda_z S^z \ , \qquad J^{x,y} = \lambda_\perp S^{x,y} \ .
\label{JSequiv}
\end{equation}
where  $S^\alpha$ are components of an $S=1/2$ pseudo-spin
defined in the subspace of the lowest doublet.
Note that the operator relations (\ref{JSequiv}) are satisfied in the local
coordinate frame for each magnetic ion.

In order to discuss possible forms of the effective Hamiltonian, it is instructive
first to assume isotropic exchange between full magnetic moments,
${\bf J}_i\cdot{\bf J}_j$, and to project the nearest-neighbor (n.n.) exchange
Hamiltonian onto the subspace of lowest CEF-doublets with the help of
Eq.~(\ref{JSequiv}) \cite{Shiba03}.
Terms with $S^z_{i(j)}$ are further suppressed by a small factor
$\lambda_z/\lambda_\perp \sim 10^{-2}$ and are neglected below.
The remaining  $XY$ interactions are generally characterized by
two parameters \cite{Curnoe08}:
\begin{equation}
\hat{\cal H} = \sum_{\langle ij\rangle}
\Bigl[\/
{\cal J} {\bf S}^\perp_i\cdot{\bf S}^\perp_j + {\cal J} _a
({\bf S}^\perp_i\cdot \hat{\bf r}_{ij})({\bf S}^\perp_j\cdot \hat{\bf r}_{ij})
\Bigr]\ ,
\label{H}
\end{equation}
where $\hat{\bf r}_{ij}= ({\bf r}_i-{\bf r}_j)/|{\bf r}_i-{\bf r}_j|$.
The first term is isotropic exchange between transverse
components of n.n.\ spins confined to the local $\{111\}$ planes.
The second term represents bond-dependent exchange anisotropy. It has
a pseudo-dipolar form but both signs of ${\cal J} _a$ are allowed by symmetry.
Although the above projection procedure yields vanishing ${\cal J}_a$,
the in-plane exchange anisotropy may be generated by virtual excitations into
higher CEF levels \cite{Shiba03}
or into states with different occupancy on rare-earth ions \cite{Onoda10}.
As we shall see below, it is essential to include
${\cal J} _a$ in the effective  Hamiltonian (\ref{H}),
since quantum fluctuations are singular in the ${\cal J}_a\to 0$ limit.

{\it Ground-state selection.}---%
The classical ground states of the Hamiltonian (\ref{H}) with ${\cal J}_a = 0$
satisfying the constraint $\sum_{\rm tet} {\bf S}^\perp_n =0$ form an infinite
but nonextensive set of states, producing planes of zero-energy modes in
the excitation spectrum \cite{Champion04}. Finite anisotropy ${\cal J}_a\neq 0$
almost eliminates this degeneracy.  For ${\cal J}_a>0$, we find that the classical
energy (\ref{H}) is minimized for a subset of $k=0$ states. Two of them, so called $\psi_2$
and $\psi_3$ states \cite{Poole07}, are shown in Fig.~\ref{fig:Mstates},
while other spin configurations are constructed as their linear combination:
\begin{equation}
{\bf S}_{n}(\varphi) =
S\bigl(\hat{\bf x}_n \cos\varphi + \hat{\bf y}_n \sin\varphi\bigr) \,,
\quad n=1\ldots 4
\label{Class}
\end{equation}
such that $\varphi=0$ $(\pi/2)$ corresponds to $\psi_2$ $(\psi_3)$,
respectively. Thus, instead of an infinite number of variables, the classical
ground-states of Eq.~(\ref{H}) for ${\cal J} _a>0$ are parameterized by one global
continuous angle, $\varphi$.

Spin states $\psi_2$ and $\psi_3$ form a basis of the two-dimensional irreducible
representation $E$ of the tetrahedral point group. They transform as
$\psi_2\sim 3z^2-r^2$ and $\psi_3\sim x^2-y^2$ under cubic rotations.
Consequently, the two spin configurations have the same mean-field energies for
any extension of the Hamiltonian (\ref{H}), involving only bilinear
spin-spin interactions ({\it e.g.}, further-neighbor exchange or
long-range dipolar interactions). The mean-field degeneracy may be lifted only by
interactions of sixth order in  spin components \cite{McClarty09},
though these are not explicitly present in the effective Hamiltonian
since $S=1/2$. Nevertheless, such interactions may be generated dynamically
by thermal or quantum fluctuations.

To study the effect of quantum fluctuations on the remaining continuous degeneracy
we use harmonic spin-wave theory. For this, spin operators in the effective
Hamiltonian (\ref{H}) are assumed to have an arbitrary spin $S$ and the usual
$1/S$ expansion is applied, substituting  $S=1/2$ in the final expressions.
This standard approach yields the following results.

Magnon excitation energies for an arbitrary classical ground-state (\ref{Class})
can be expressed as
\begin{equation}
\varepsilon^{(n)}_{\bf k} = {\cal J}S\sqrt{(2+j_a)^2+2(2+j_a)\lambda^{(n)}_{\bf k}} \ ,
\label{Ek}
\end{equation}
where, $j_a = {\cal J}_a/{\cal J}$ and $\lambda^{(n)}_{\bf k}$ are four eigenvalues of
the matrix $C_{\bf k}$,
\begin{equation}
\label{Mk}
C_{\bf k} =\left(\begin{array}{cccc}
0        &  c^+_{xy} & c^+_{yz} & c^+_{xz} \\[0.4mm]
c^+_{xy} &  0        & c^-_{xz} & c^-_{yz} \\[0.4mm]
c^+_{yz} &  c^-_{xz} & 0 & c^-_{xy} \\[0.4mm]
c^+_{xz} &  c^-_{yz} & c^-_{xy}   &  0
\end{array}\right)
\end{equation}
with $c^\pm_{xy} = t_{xy} \cos(k_x\pm k_y)/4$, \ldots, and
\begin{eqnarray}
& &
t_{xy} = \frac{4}{3}\sin(\varphi-\pi/3)\sin(\varphi+\pi/3) - \frac{j_a}{3}\sin^2\!\varphi \,,
 \\
& &
t_{yz} =-\frac{4}{3}\sin\varphi \sin(\varphi-\pi/3) - \frac{j_a}{3}\sin^2(\varphi+\pi/3) \,,
\nonumber \\
& &
t_{xz} =-\frac{4}{3}\sin\varphi \sin(\varphi+\pi/3) - \frac{j_a}{3}\sin^2(\varphi-\pi/3) \,.
\nonumber
\end{eqnarray}

\begin{figure}[t]
\centerline{
\includegraphics[width=0.8\columnwidth]{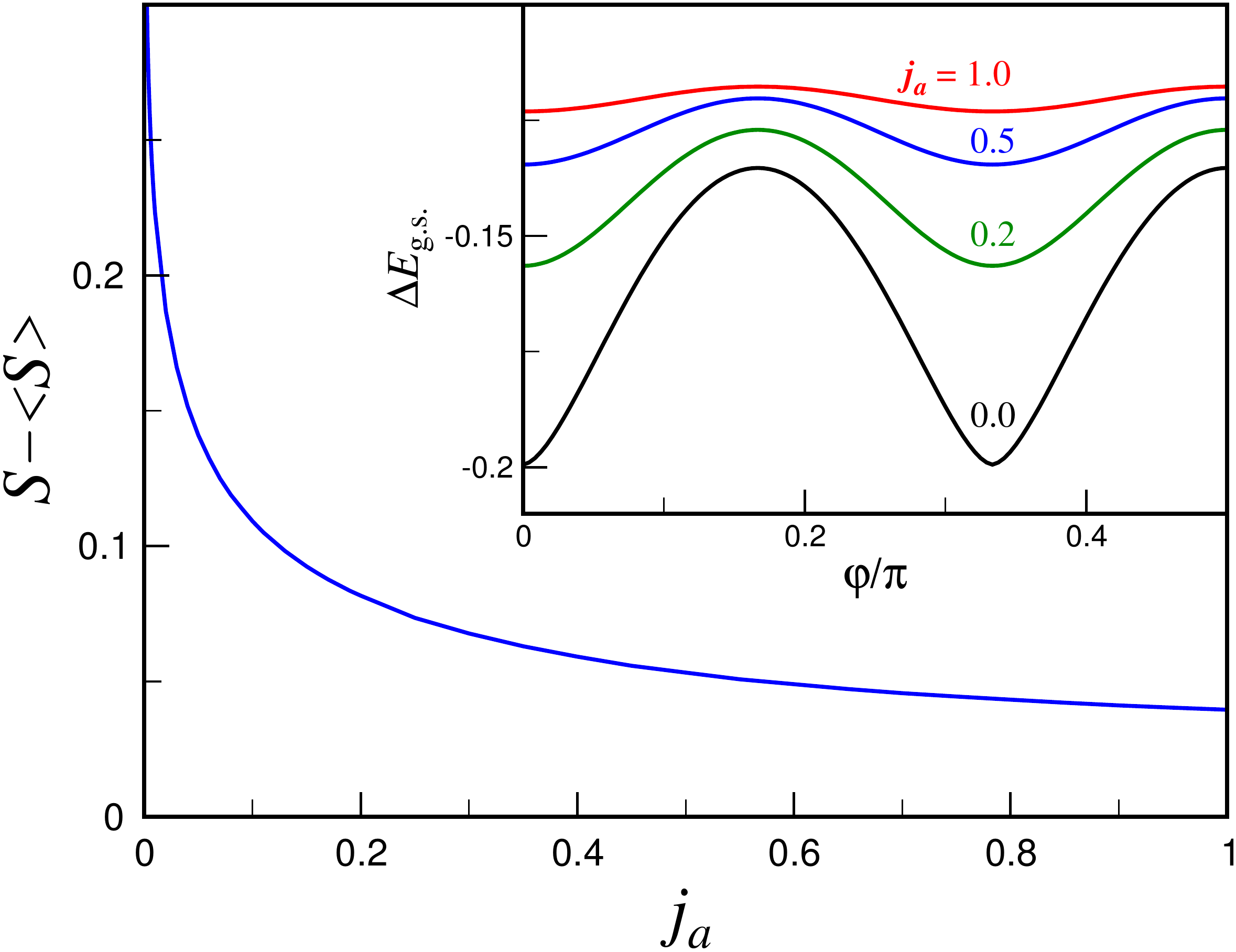}}
\caption{(Color online) Inset: the ground-state energy in units of $\cal J$
for degenerate classical configurations parameterized by angle $\varphi$,
see Eq.~(\ref{Class}). Main panel: the quantum spin reduction
in the noncoplanar state $\psi_2$ versus the bond-dependent exchange anisotropy
parameter $j_a ={\cal J}_a/{\cal J}$.
}
\label{fig:SWT}
\end{figure}

The quantum correction to the ground-state energy normalized per spin is
\begin{equation}
\Delta E_{\rm g.s.} = \frac{1}{8}\sum_{n=1}^4 \int\frac{\mathrm{d}^3k}{V_{\rm BZ}}\,
\varepsilon^{(n)}_{\bf k} - \frac{1}{2}{\cal J}S(2+j_a) \ ,
\label{Egs}
\end{equation}
where $V_{\rm BZ}$ is the  Brillouin zone volume. $\Delta E_{\rm g.s.} (\varphi)$
is shown in the inset of Fig.~\ref{fig:SWT} for $S=1/2$, exhibiting a clear minimum
for the noncoplanar $\psi_2$ state ($\varphi=0,\pi/3,\ldots$) for any value of $j_a$.
Hence quantum fluctuations, like thermal fluctuations, select the $\psi_2$ state
from the ensemble of classical ground states giving an example of quantum order by disorder.
A similar calculation for isotropic exchange $j_a=0$, was recently presented in
\cite{Stasiak11}. However, as shown below, quantum fluctuations become
singular and essentially nonperturbative in the isotropic case and may completely
destroy the long-range magnetic order.

Harmonic spin-wave theory also yields the leading quantum correction to the value
of the ordered moment $\langle S\rangle$. Most conveniently this can be done with
the help of a fictitious staggered field $h_s\to 0$. The corresponding expression
for $\delta S = S- \langle S\rangle$  is
\begin{equation}
\delta S =  \frac{1}{8} \sum_{n=1}^4\! \int\!\frac{\mathrm{d}^3k}{V_{\rm BZ}}\,
\frac{2+j_a+\lambda^{(n)}_{\bf k}}
{\sqrt{(2+j_a)^2+2(2+j_a)\lambda^{(n)}_{\bf k}}} - \frac{1}{2} \,.
\label{dS}
\end{equation}

The spin reduction $\delta S$ in the ordered $\psi_2$ state (Fig.~\ref{fig:SWT})
diverges logarithmically as $j_a\to 0$. This is a result of the effective dimensional
reduction to one dimensional behavior, due to the planes of zero-energy modes
in the magnon dispersion $\varepsilon_{\bf k}$ in the isotropic case  \cite{Champion04}.
However, a small anisotropy of $j_a \simeq 0.05$ already cuts off the singularity
and restores the perturbative nature of the quantum correction. The measured
ordered moment of
$\mu=3.2\mu_B$ in $\rm Er_2Ti_2O_7$ \cite{Poole07} indicates a 16\% renormalization
from the nominal value $\mu_0\approx 3.8\mu_B$ for the Er$^{3+}$ ground-state
doublet \cite{Champion03}.
This gives an estimate of  $j_a\sim 0.3$--0.5. The large uncertainty is related
to the unknown precision for $\mu_0$, which cannot be measured directly,
but our analysis is consistent with a perturbation of considerable size.

{\it Excitation spectrum.}---%
With both Hamiltonian and its ground state in hand, we compute the spin-wave spectrum
in $\rm Er_2Ti_2O_7$. The $\psi_2$ magnetic structure allows six different domains
with $\varphi =\pi\,n/3$. We choose $\varphi=0$ (pictured in Fig.~\ref{fig:Mstates}(a))
with spins pointing along $[\pm 1,\!\pm 1,\!\pm 2]$ cubic directions. Analytical
expressions for magnon energies can be derived from Eq.~(\ref{Ek}) for a few symmetry
directions in the Brillouin zone. For ${\bf k}=(q,0,0)$, the four branches are
\begin{equation}
\Bigl(\frac{\varepsilon_{\bf k}}{{\cal J}S}\Bigr)^2 \!=\!
(2+j_a)\Bigl[2+j_a\mp\, 2\cos\frac{q}{4}
\pm\,{\textstyle \frac{1}{2}}j_a\bigl(1+\cos\frac{q}{4}\bigr)\Bigr].
\label{E100}
\end{equation}
Along the ${\bf k}=(0,0,q)$ line, the two dispersive branches are
expressed as
\begin{equation}
\varepsilon_{\bf k} = {\cal J}S\sqrt{(2+j_a)j_a[1\pm \cos(q/4)]}\ ,
\label{E001}
\end{equation}
while the two other modes  remain dispersionless with
$\varepsilon_{\bf k} = {\cal J}S\sqrt{(2+j_a)(4+j_a)}$.

The characteristic low-energy feature of the calculated spectra
is a gapless mode with acoustic dispersion
$\varepsilon_{\bf k}\approx ck$ for $k\to 0$.
This remarkably
reproduces  a variety of experimental results for $\rm Er_2Ti_2O_7$
including the $T^3$-dependence of the heat capacity \cite{Blote69},
the inelastic neutron data \cite{Ruff08},
and the electron spin resonance measurements \cite{Sosin10}.
The presence of the gapless branch in the harmonic approximation
is directly linked to the classical degeneracy of the model (\ref{H})
and is, therefore, accidental in the sense that is
not related to any symmetry of the magnetic system.
Hence, a finite gap will generically be produced by quantum fluctuations
in higher orders of the spin-wave theory. Corresponding calculations
are rather cumbersome and are not attempted here. We only remark
that the generated quantum gap should depend on the curvature of
$\Delta E_{\rm g.s.}(\varphi)$ in the vicinity of its minimum.
Since the minimum becomes more shallow with increasing anisotropy,
see the inset of Fig.~\ref{fig:SWT}, we expect that for $j_a\geq 0.5$
the actual gap may be rather small.

\begin{figure}[b]
\centerline{
\includegraphics[height=0.6\columnwidth]{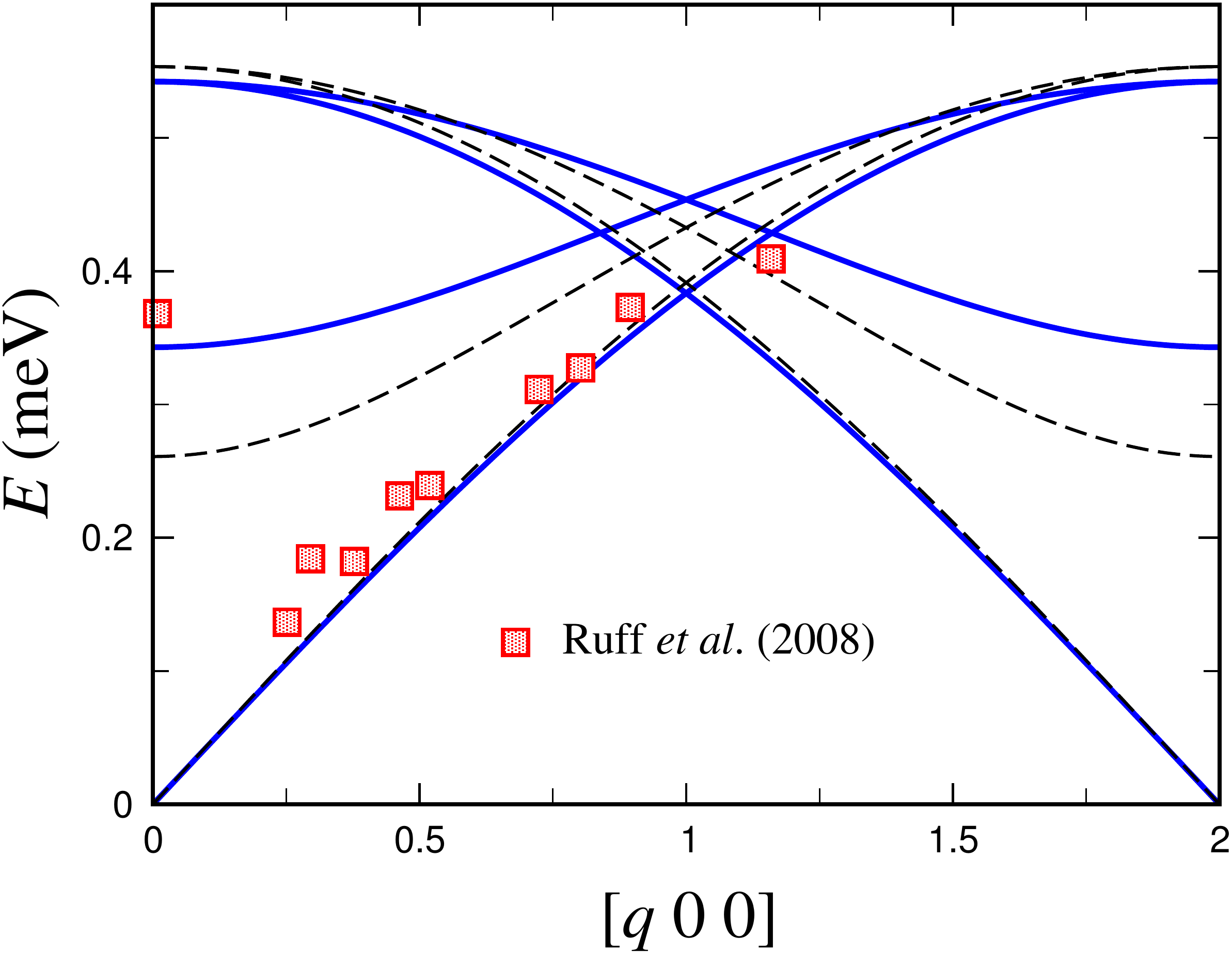}  }
\caption{(Color online) Zero-field dispersion of magnons in $\rm Er_2Ti_2O_7$
along the $[1 0 0]$ direction. Wave-vectors are measured in units of $2\pi$.
Solid lines are calculated for ${\cal J}=0.28$~meV and $j_a=1$, while
dashed lines are for ${\cal J}=0.33$~meV and $j_a=0.5$.
Open squares are experimental data points \cite{Ruff08}.
}
\label{fig:spectrum}
\end{figure}

Figure~\ref{fig:spectrum} shows the dispersion of four magnon branches
in the $[1 0 0]$ direction in comparison with the inelastic neutron scattering
data of Ruff {\it et al.}\ \cite{Ruff08} scanned from their Fig.~4.
We find particularly good agreement  between theory and experiment for the
following values of the microscopic parameters:
${\cal J}=0.28$~meV and $j_a=1$, corresponding curves are plotted by
solid lines. Magnon energies for
${\cal J}=0.33$~meV and $j_a=0.5$ a value extracted from the reduced moment (see above) are shown by dashed lines.
While the low-energy branch is fitted equally well, a larger $j_a$ is favored by the position of
the lower gap $\Delta\approx 0.38$~meV.

We can extract further information about the the microscopic parameters
from values of the critical field between the antiferromagnetic
phase and the polarized paramagnetic state \cite{Ruff08,Cao10,Sosin10,Petrenko11}.
The high-field polarized states are particularly
simple for field orientations along [100] and [110] axes.
Repeating the above spin-wave calculations for ${\bf H}\parallel[110]$ we obtain
the critical field from the vanishing gap in the magnon spectrum
\begin{equation}
g_\perp\mu_B H_c^{[110]} =
{\cal J}S\Bigl[4 + \frac{4}{\sqrt{3}} + j_a\bigl(\frac{2}{\sqrt{3}}-1\bigr) \Bigr] \ ,
\label{Hc_110}
\end{equation}
where $g_\perp = g_J\lambda_\perp\approx 7.6$. The dependence of (\ref{Hc_110})
on $j_a$ is very weak and from the extrapolated $T=0$ value
$H^{[110]}_c\simeq 1.8$~T \cite{Ruff08,Cao10,Sosin10},  we derive
${\cal J}\simeq 0.25$~meV, which is consistent with the estimate based on
neutron data. Additional measurements of the excitation spectrum and/or
the angular dependence of the critical field $H_c$ at low temperatures
should allow a more precise identification of the microscopic parameters of
the model (\ref{H}) in the case of $\rm Er_2Ti_2O_7$.

\begin{figure}[t]
\centerline{
\includegraphics[width=0.8\columnwidth]{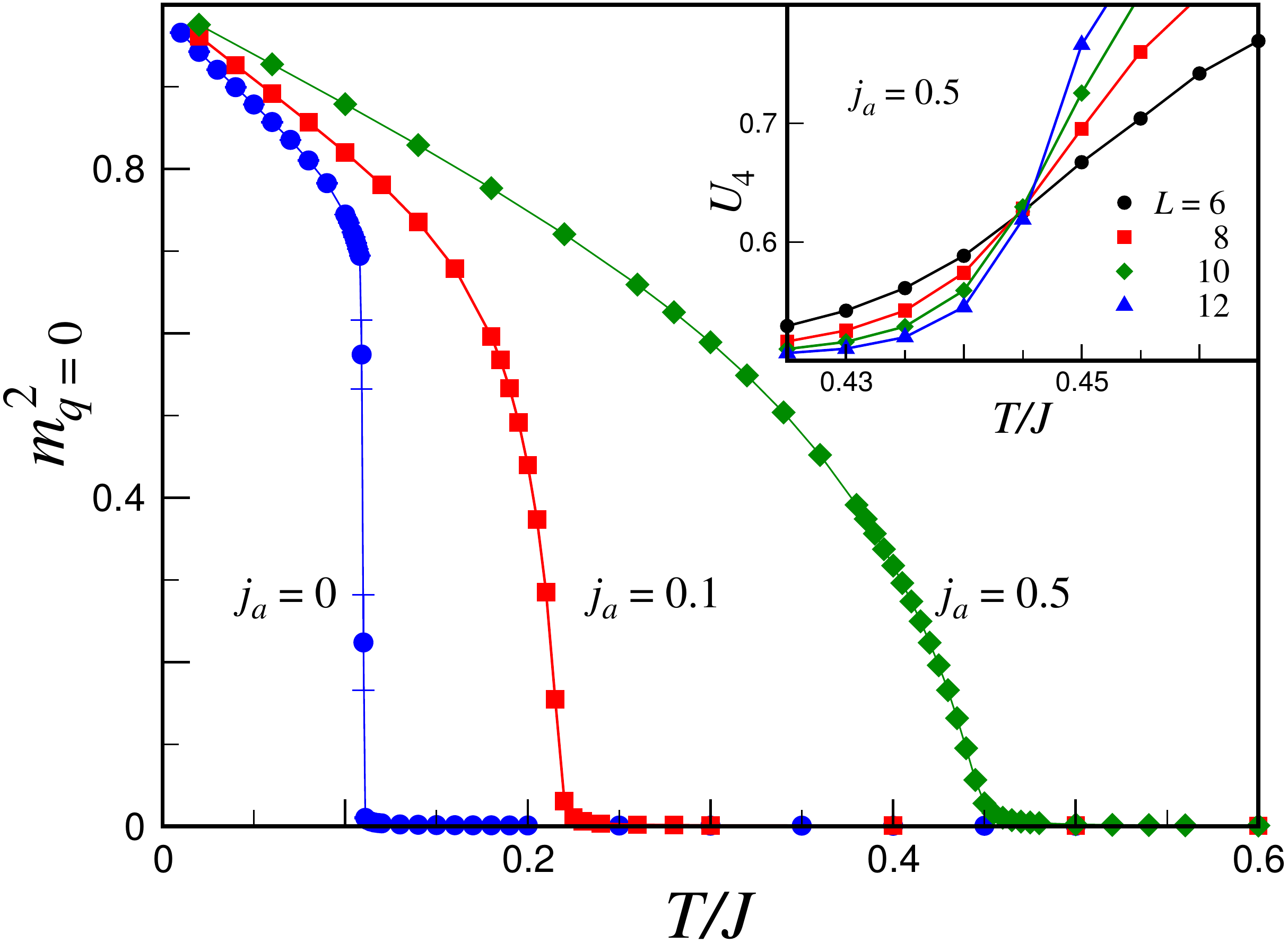}}
\caption{(Color online) Main panel: temperature dependence of the sublattice
magnetization for the classical $XY$ pyrochlore antiferromagnet with in-plane
anisotropic exchange on periodic cluster with $L=12$. Inset: the fourth-order
Binder cumulant for several lattice sizes at fixed $j_a  = 0.5$.
}
\label{fig:subl}
\end{figure}

{\it Finite-temperature transition.}---%
We finally consider the nature of ordering transition
in the $XY$ pyrochlore antiferromagnet with the bond-dependent
exchange anisotropy. Since the original quantum model (\ref{H})
remains intractable for numerical investigation, we study instead
a simplified version of (\ref{H}) with spins represented by classical
unit  vectors. The classical Monte Carlo simulations were performed on
periodic lattices of $N=4L^3$ spins with linear sizes $L=6$--12 using
the Metropolis algorithm with $\sim 10^7$ Monte Carlo steps.
Temperature dependence of the sublattice magnetization $m_{q=0}$ for
antiferromagnetic ordering with zero wave-vector is shown
in Fig.~\ref{fig:subl} for three values of $j_a$. For the model
with isotropic exchange, $j_a=0$, a clear jump appears at $T_c/J = 0.110(1)$.
However, already a small exchange anisotropy
$j_a=0.1$ changes the behavior $m_{q=0}(T)$ to continuous.

This is confirmed by the temperature dependence of the fourth-order Binder cumulant
\begin{equation}
U_4 = \frac{\langle m_{q=0}^4\rangle } {\langle m_{q=0}^2\rangle^2 } \ .
\end{equation}
According to finite-size scaling theory, Binder cumulants for clusters
with different linear sizes $L$ cross at the critical point where spin-spin
correlations exhibit a power-law decay. The crossing point $T_c/J = 0.445(2)$
for $j_a=0.5$ (Fig.~\ref{fig:subl}) gives an accurate estimate for the
transition point in this case. In our Monte Carlo simulations we also
investigated various Potts-type order parameters, which distinguish
between the noncoplanar $\psi_2$ and the planar $\psi_3$ spin structure.
As $T\to 0$ there is a clear selection of the $\psi_2$ configuration.
However, at higher temperatures spins strongly fluctuate
restoring at $T\to T_c$ the continuous $XY$
degeneracy parameterized by angle $\varphi$ in (\ref{Class}).

In conclusion, we have proposed a simple quantum model for the $XY$
pyrochlore antiferromagnet  $\rm Er_2Ti_2O_7$, which includes bond-dependent
in-plane exchange anisotropy of strength $j_a$. We show that the order
by disorder selection occurring for $j_a=0$ is maintained at both quantum
and classical level for perturbations up to at least the strength of
the isotropic exchange. The exchange anisotropy also reproduces the observed second-order transition
found in $\rm Er_2Ti_2O_7$ in thermodynamic measurements.
Further work is planned to investigate field evolution
of the antiferromagnetic state and role of the long-range part of the dipolar
interactions, but it is worth noting here that the perturbation is of the same form,
{\em but of opposite sign} to the dipolar interaction for n.n.\ pairs of spins.
As noted by Stasiak {\it et al}.~\cite{Stasiak11}, 
an exchange anisotropy of this kind can therefore overcome 
the n.n.\ part of the dipolar interaction. In addition, the long-range part
of the dipolar interaction
loses importance as the tetrahedral units have
no net magnetic moment in the ground state. These observations provide
a possible explanation as to why the dipolar effects appear to be absent
in $\rm Er_2Ti_2O_7$.

The fact that our theoretical analysis with single parameter
$j_a$ is in agreement with many experimental features
of  $\rm Er_2Ti_2O_7$ strongly supports the above picture and in the process
provides strong evidence that the ordering transition and the stability of the
experimentally observed noncoplanar magnetic structure
is indeed a remarkable experimental realization of
the order by disorder mechanism \cite{Champion03}.
A consequence of this is the unique appearance of a non-Goldstone acoustic magnon branch
with a very small gap.

We are grateful to Z. Hao and S. Sosin for their help at the initial stage of this project.
We thank S. Bramwell, P. Dalmas de R\'eotier, B. Gaulin, M. Gingras, and S. Sosin for many fruitful
discussions.  MEZ and PCWH acknowledge hospitality of the Max Planck Institute for the
Physics of Complex Systems, where part of this work has been done.

{\it Note added in proof.}---After submission of this work we became aware of a
related preprint by Savary {\it et al}.\ \cite{Savary12b}.


\end{document}